\documentclass[lettersize,journal]{IEEEtran}
\usepackage{amsmath,amsfonts}
\usepackage{algorithmic}
\usepackage{algorithm}
\usepackage{array}
\usepackage[caption=false,font=normalsize,labelfont=sf,textfont=sf]{subfig}
\usepackage{textcomp}
\usepackage{stfloats}
\usepackage{bibentry}
\usepackage{url}
\usepackage{verbatim}
\usepackage{graphicx}
\usepackage{cite}
\usepackage{multirow}
\usepackage{balance}
\usepackage{makecell}
\usepackage{stfloats}
\usepackage{amssymb}    
\usepackage{threeparttable}
\usepackage{upgreek}

\def\BibTeX{{\rm B\kern-.05em{\sc i\kern-.025em b}\kern-.08em
		T\kern-.1667em\lower.7ex\hbox{E}\kern-.125emX}}
\usepackage{balance}
\begin{document}
	\title{ Measurement-Based Non-Stationary Markov  Tapped Delay Line Channel Model for 5G-Railways}
	\author{Xuejian Zhang, ~\IEEEmembership{Student Member,~IEEE,} 
	Ruisi He,~\IEEEmembership{Senior Member,~IEEE,} 
	Mi Yang,~\IEEEmembership{Member,~IEEE,} \\
	Jianwen Ding,~\IEEEmembership{Member,~IEEE,}
	Ruifeng Chen,
	Shuaiqi Gao, Ziyi Qi, Zhengyu Zhang,~\IEEEmembership{Student Member,~IEEE,} \\
	Bo Ai,~\IEEEmembership{Fellow,~IEEE, }	
	and Zhangdui Zhong,~\IEEEmembership{Fellow,~IEEE}
	\thanks{

X. Zhang, R. He,  J. Ding, S. Gao, Z. Qi, Z. Zhang,  B. Ai, and Z. Zhong are with the School of Electronics and Information Engineering and the Frontiers Science Center for Smart High-speed Railway System, Beijing Jiaotong University, Beijing 100044, China
(email: 23115029@bjtu.edu.cn; ruisi.he@bjtu.edu.cn;  jwding@bjtu.edu.cn; 23120047@bjtu.edu.cn; 22115006@bjtu.edu.cn; 21111040@bjtu.edu.cn; boai@bjtu.edu.cn; zhdzhong@bjtu.edu.cn).

M. Yang is with the School of Electronics and Information Engineering and the Frontiers Science Center for Smart High-speed Railway System, Beijing Jiaotong University, Beijing 100044, China, and also with the Henan High-Speed Railway Operation and Maintenance Engineering Research Center, Zhengzhou 451460, China (e-mail: myang@bjtu.edu.cn).

R. Chen is with the Institute of Computing Technology, China
Academy of Railway Sciences Corporation Ltd., Beijing 100081, China
(e-mail: ruifeng\_chen@126.com).

}
}
	\maketitle

\begin{abstract}
5G for Railways (5G-R)  is globally recognized as a promising next-generation railway communication system designed to meet   increasing demands.
Channel modeling serves as   foundation for communication system design, with tapped delay line (TDL) models widely utilized in system simulations due to their simplicity and practicality and serves as a crucial component of various standards like 3GPP.
However, existing TDL models applicable to 5G-R systems are limited. 
Most fail to capture  non-stationarity, a critical characteristic of railway communications, while others are unsuitable for the specific frequency bands and bandwidths of 5G-R.
In this paper, a  channel measurement campaign for  5G-R dedicated network is carried out, resulting in   a measurement-based 5-tap TDL model utilizing a first-order two-state Markov chain to represent channel non-stationarity. Key model parameters, including  number of taps, statistical distribution of amplitude, phase and Doppler shift, and state transition probability matrix, are extracted. 
The correlation between tap amplitudes are also obtained. 
Finally,  accuracy of model is validated through comparisons with measurement data and 3GPP model.
These findings are expected to offer valuable insights for  design, optimization, and link-level simulation and validation of 5G-R systems.

\end{abstract}

\begin{IEEEkeywords}
	5G-R, channel measurement, TDL model, non-stationarity, Markov chain.
\end{IEEEkeywords}

\section{Introduction}
\IEEEPARstart{I}{n} modern society, railway transportation has become an indispensable mode of transit for both passenger travel and goods transport \cite{ber2021}. 
Dedicated railway mobile communication systems play a crucial role in ensuring the safety, efficiency, and reliability of railway operations \cite{zemen2024}. 
Over the past few decades,  the Global System for Mobile Communications for Railway (GSM-R) has been widely deployed worldwide due to its exceptional reliability \cite{he2016}. 
However, with the rapid proliferation of 5G technology and infrastructure in public networks, railway communication systems face heightened demands, including requirements for higher bandwidth, higher data transmission rate, and enhanced multimedia transmission capabilities \cite{zt2019-2}. 
Unfortunately, GSM-R is no longer sufficient to meet these emerging needs. Consequently, countries such as China and those in Europe are actively pursuing the modernization and evolution of dedicated railway networks by incorporating 5G technology into railway systems, namely  5G for Railways (5G-R) \cite{he2022}.
In 2023, China authorized the 5G-R test frequency band: 1965–1975 MHz for uplink and 2155–2165 MHz for downlink \cite{liang2024}. 
Similarly, Europe is advancing the Future Railway Mobile Communication System (FRMCS), which proposes 5GRAIL \cite{5GRail} as the successor to GSM-R, operating within an authorized frequency band of 1900–1910 MHz.
The allocation of dedicated frequency bands for  5G-R system has provided clear guidance for research on air interface technologies and  channel characterization \cite{he2023}.

Accurate channel modeling is foundational to design, optimization, and deployment of wireless communication systems, particularly for railway communication systems \cite{zxj2023,he2024-2}. 
Numerous studies have been conducted on channel modeling in railway communications. 
For instance, \cite{hrs2011} discusses  the models of path loss and shadow fading  in viaduct scenarios at 930 MHz.  
Wideband channel characteristics, such as root mean square delay spread (RMS DS), power delay  profile (PDP), and angular spread, are analyzed in \cite{liu2012, zt2019}, covering frequencies from 1.89 GHz to 2.605 GHz. 
Besides  typical channel parameters, 
link-level models, such as TDL models, are also important in channel modeling. 
As a key component of standards like 3GPP \cite{3gpp} and WINNER II \cite{winner}, TDL model is valued for its simplicity and robust performance. 
It can effectively guide link-level simulations across various scenarios and frequency bands and has been proposed for railway communications as well \cite{he2024}. 
At 950 MHz and 2150 MHz with a bandwidth of 100 MHz, \cite{ding2017} establishes  2–6 taps TDL models based on measurements from composite scenarios of cuttings and tunnels. 
\cite{qiu2021, yjy2019}   utilize ray-tracing simulation datasets to develop various TDL models for railway tunnels and cuttings, covering frequencies between 2.15 GHz and 2.4 GHz. 
Additionally, measurement-based TDL models are established in hilly areas and viaduct scenarios at 2.4 GHz and 2.6 GHz with a bandwidth of 40 MHz \cite{zy2014, qian2014}.

Although TDL models have been established in various railway scenarios, none of the aforementioned models can describe the fast time-varying or non-stationarity of   channel.
They are all based on the idealized Wide-Sense Stationary Uncorrelated Scattering (WSSUS) assumption, which is also adopted by standards such as 3GPP \cite{3gpp}.
However, for real high-speed railways, trains move at high speeds, such as 350 km/h, and the surrounding scatterers change dynamically, rendering the WSSUS condition invalid \cite{liu2018, zt2019-2, wu2018}. 
Unfortunately, none of the above studies addresses channel modeling under the non-WSSUS conditions.

To our knowledge, only a limited number of studies have explored these challenges. TDL models based on first-order and second-order Markov chains for viaduct scenarios at 2.35 GHz are proposed in \cite{liul2014,liu2012} , using state transition probability matrix to describe the ``birth" and ``death"  of multipath components (MPCs). 
This approach is widely adopted in vehicular communications to establish non-stationary TDL models \cite{sen, hassan, ly2013,cje}. 
However, more studies on non-stationary TDL models in 5G-R frequency bands and bandwidth cannot be found. 
This gap significantly hinders the implementation of link-level simulations and  performance evaluations (e.g., channel capacity and throughput) for 5G-R systems.

To address this research gap, we conduct  comprehensive channel measurement campaigns for 5G-R private networks in this paper, capturing extensive wideband channel data. Based on  measured data, a 5-tap TDL model is developed using a first-order two-state Markov chain to characterize the persistence of MPCs and  non-stationarity of wireless channel. 
Finally, the proposed model is   compared with the measured data and 3GPP model to validate its accuracy, providing a robust foundation for the future development of 5G-R systems.

The rest of the paper is organized as follows. 
The measurement campaign is introduced in Section II. 
In Section III, a Markov-based TDL model is proposed and validated. 
Afterwards, Section IV draws conclusions.

\section{Measurement Campaign}
\subsection{Measurement Scenario}
\begin{figure}[!t]
	\centering
	\includegraphics[width=.45\textwidth]{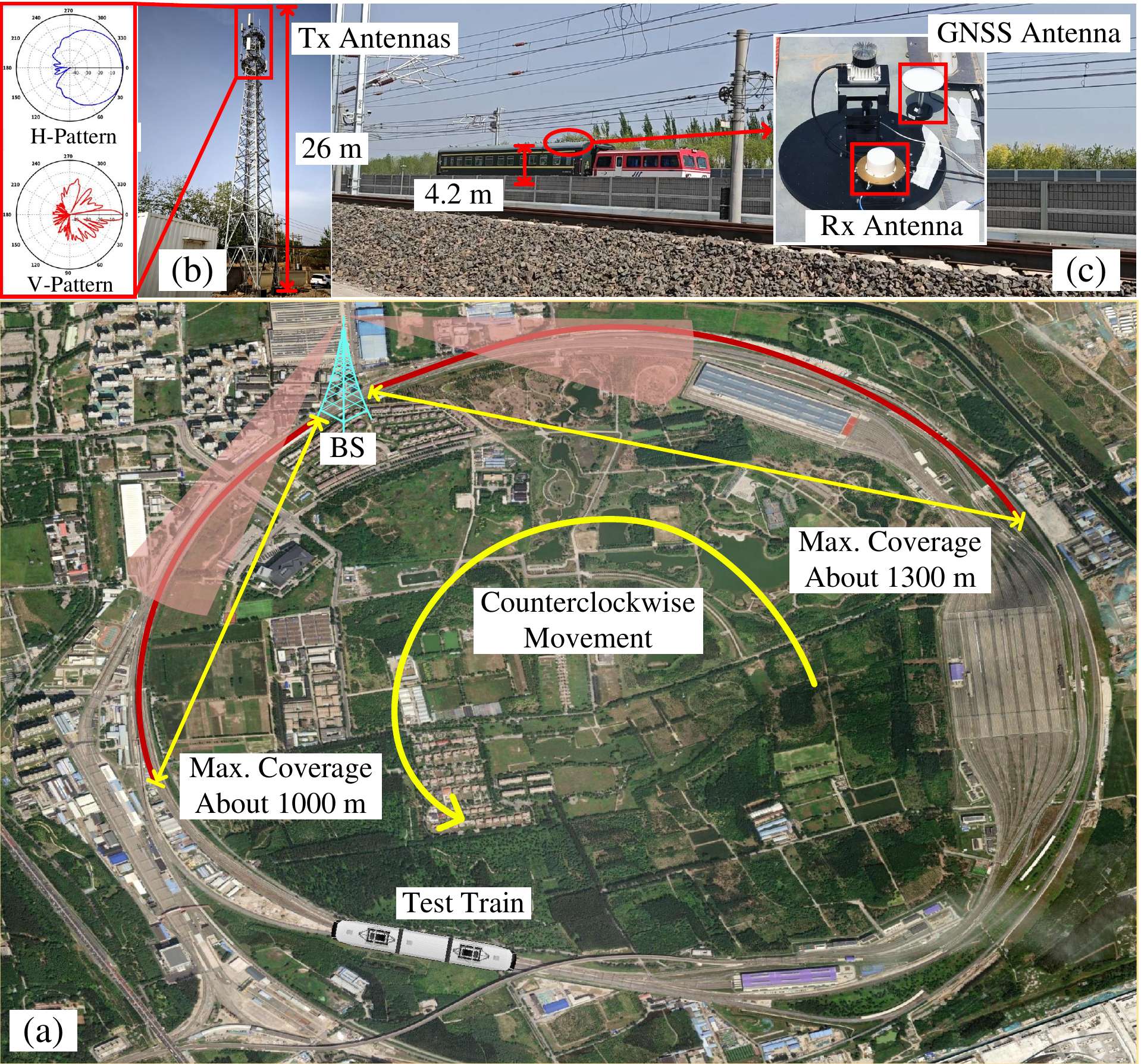}
	\caption{Measurement system and scenario.}
	\label{scenario}
\end{figure}

A comprehensive 5G-R channel measurement campaign is conducted at the National Railway Track Test Center, situated northeast of Beijing, China, as illustrated in Fig. \ref{scenario}(a). 
Along this railway,  a 5G-R dedicated  network base station (BS) is deployed, with two antennas radiating in opposite directions.
To mitigate signal interference from the public network and neighboring stations, the frequency band of 2155–2165 MHz is cleared. 
The measured scenario is classified as a rural area, predominantly surrounded by trees and scattered low-rise buildings. 
Furthermore, specific railway-related structures, such as low partition walls and contact network poles along the railway, contribute additional unique scatterers within the measurement environment.

Transmitting (Tx) antennas are mounted on the tower, as shown in Fig. \ref{scenario}(b), and  connected to a vector signal generator and power amplifier under the tower, respectively.
The receiving (Rx) system consisted of a vector signal analyzer positioned inside the cabin of  test train, with a Rx antenna mounted on its roof as shown in Fig. \ref{scenario}(c).
Tx antennas are $\pm\ 45^{\circ}$ polarized directional panel antenna with 17.5 dBi gain, while Rx antenna is a vertically polarized omnidirectional antenna with 3 dBi gain. 
To achieve precise time synchronization, two rubidium clocks tamed by Global Navigation Satellite System (GNSS) signals are  employed.
A frequency of 2.16 GHz with bandwidth of 10 MHz is used during the measurement campaign, which is consistent with the 5G-R dedicated test frequency band allocated by China, resulting in a resolution delay of $\Delta \tau = 100 $ ns. 
The test train maintain  a constant speed of 80 km/h  as shown in Fig. \ref{scenario}(c). 
It travels counterclockwise along the circular railway track for multiple laps to ensure the collection of sufficient channel data. 

%

\subsection{Data Processing}
Following the acquisition of  raw measured data, the first step is to perform data calibration. 
This process includes back-to-back  and antenna calibration to mitigate the effects of cables, transceivers, and antenna radiation patterns, etc \cite{zxj2024}.
Subsequently, Channel impulse response (CIR)  is derived, denoted as $ h(t,\tau ) $ where $t$ is  index of measurement time snapshot, $ \tau $ is   delay bin. 
It is accomplished by first calculating the channel transfer function in the frequency domain, incorporating system calibration, and then applying an inverse Fourier transform.
A detailed description of calibration and processing methodology can be found in \cite{yangmi2019}.

\section{Markov-based TDL model for 5G-R }
\subsection{Markov-Based TDL Model} 
In the TDL model, a set of discrete paths is defined, each characterized by a specific delay and attenuation level. 
Additionally, each path is associated with a Doppler spectrum to account for channel variability caused by the movement of Rx. 
What's more, a notable characteristic of railway channels is the invalidity of WSSUS assumption, as the propagation environment changes rapidly \cite{liu2014}. 
Scatterers that vary rapidly over time cause MPCs to  undergo a "birth and death" process, indicating the non-stationarity   of  railway channel \cite{zhangbei}.
Then, the TDL model can be expressed as
\begin{equation}
	\label{math1}
	h(\tau ,t) = \sum\limits_{l = 1}^L {{z_l}(t){\alpha _l}(t){e^{j\left[ {{\phi _l}(t) + 2\pi {f_{D,l}}(t)} \right]}}} \delta \left[ {\tau  - {\tau _l}(t)} \right],
\end{equation}
where ${\alpha _l}(t)$, ${\phi _l}(t)$, ${\tau _l}(t)$ and ${f_{D,l}}$ is the amplitude, phase, time delay, and Doppler shift of the $l$th resolvable tap, respectively. 
Additionally, $\delta(\cdot)$ denotes the Dirac delta function, and $L$ represents the number of taps.
Compared to WSSUS TDL models, a switching function ${z_l}(t) = {0, 1}$ is proposed to account for the finite ``lifetime" of   MPCs, which reflects the non-stationarity of railway channels.
This phenomenon occurs at a rate slower than small-scale fading (such as multipath or Doppler effects) but faster than large-scale effects (such as path loss or shadowing), representing the ``birth" or ``death" of MPC over short time or spatial intervals \cite{liul2014,liu2012}.
Therefore, the behavior of ${z_l}(t)$ can be characterized as a non-deterministic process such as Markov chain.
Specifically, we use  the first-order two-state Markov chain to characterize the ``birth" and ``death" of   MPCs as \cite{zhangbei}
\begin{equation}
	\label{math2}
	T = \left[ {\begin{array}{*{20}{c}}
			{{p_{00}}}&{{p_{01}}}\\
			{{p_{10}}}&{{p_{11}}}
	\end{array}} \right],S = \left[ {\begin{array}{*{20}{c}}
			{{p_0}}\\
			{{p_1}}
	\end{array}} \right],
\end{equation}
where $T$ is the state transition matrix, $p_{mn}$ is the transition probability of transition from state $m$ to state $n$, the state 0 represents the ``death" of taps, and state 1 represents the ``birth" of taps. 
$S$ is the steady-state matrix and $p_{w}$ means the steady-state probability of state $w$, which satisfies ${p_{00}} + {p_{01}} = 1$, ${p_{10}} + {p_{11}} = 1$, and ${p_{0}} + {p_{1}} = 1$. 
The state decision interval of ``birth and death" state is defined as channel coherence time. 
      

\begin{figure}[!t]
	\centering
	
	\subfloat[]{\includegraphics[width=.24\textwidth]{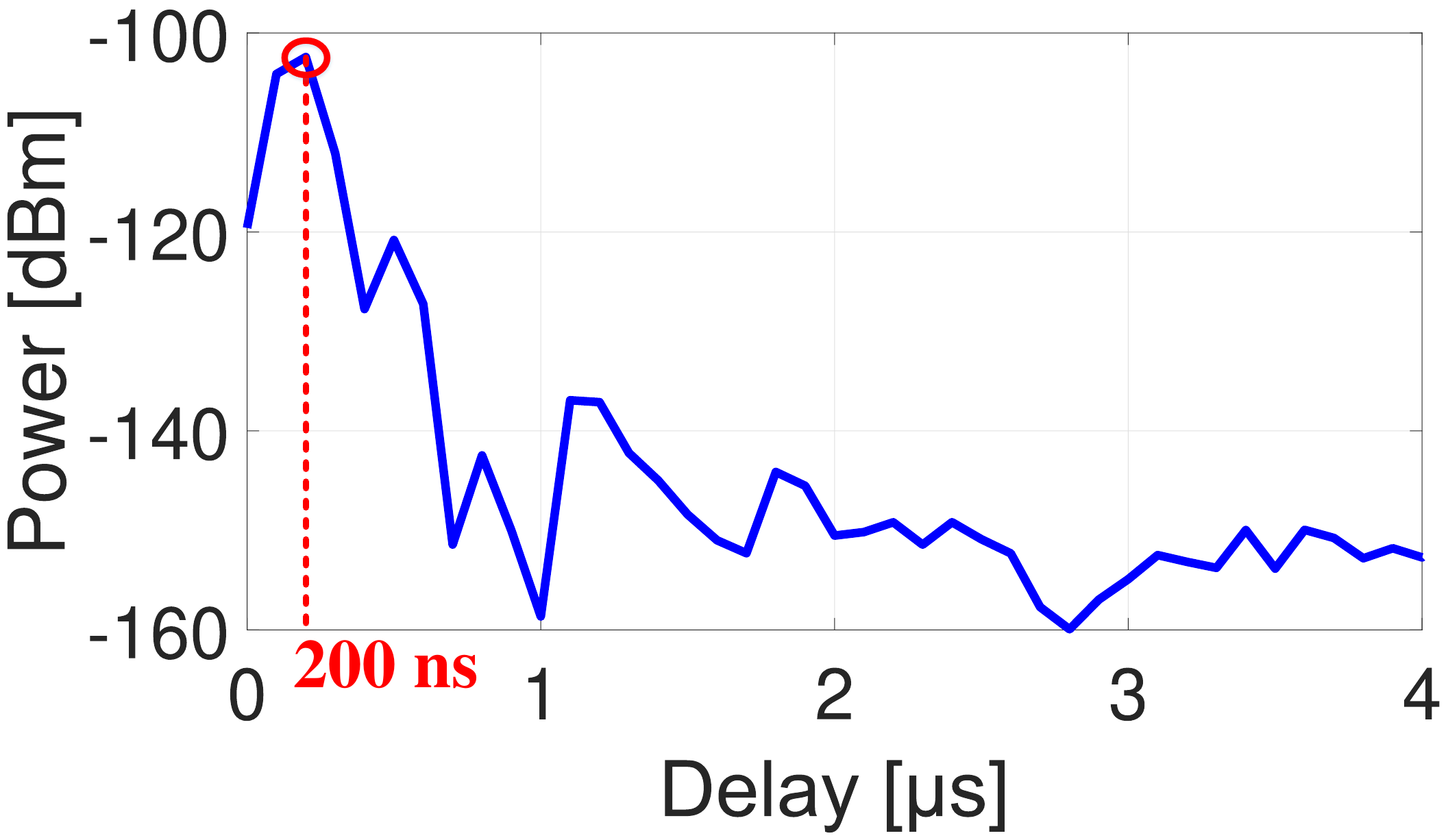}%
		\label{amplitude}}
	\subfloat[]{\includegraphics[width=.24\textwidth]{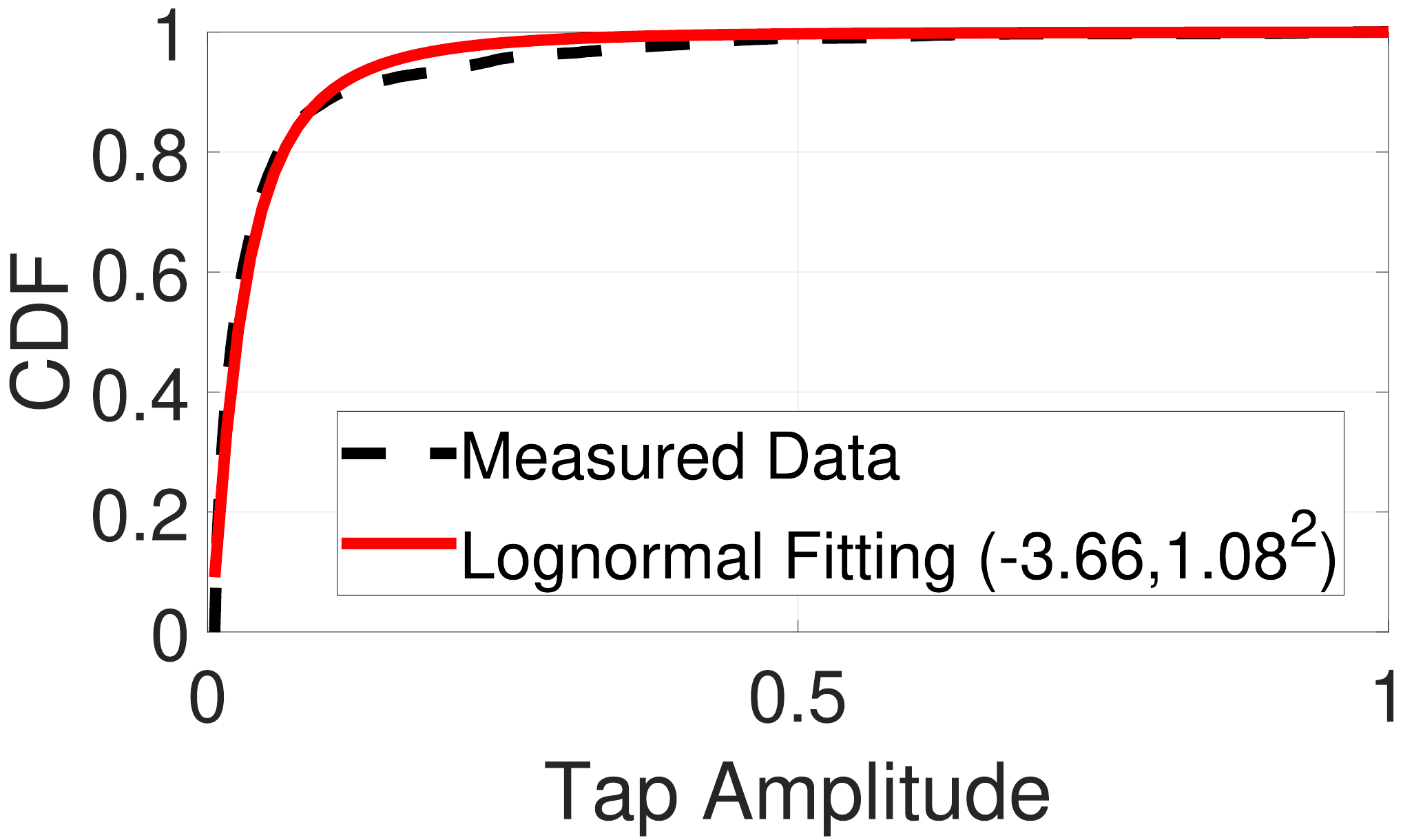}%
		\label{PDP}}
	\caption{
		(a) An example of PDP  in one snapshot.
		(b) CDFs of the Lognormal fitting of   tap amplitudes.
	}
	\label{amplitute}
\end{figure}

\begin{table}
	\renewcommand{\arraystretch}{1.2} 
	\begin{center}
		\caption{Parameters of Markov-Based TDL Model .}
		\label{TDL}
		\begin{threeparttable}
			\begin{tabular}{ c| c| c| c|   c| c }
				\hline
				\hline
				Tap  & \makecell[c]{Delay  [$\upmu$s] }  &Power  [dB]  & $P_{00}$ &  $P_{11}$ & $P_{1}$   \\
				\hline
				\makecell[c]{1 }   & 0 & 0 & 0 &  1 & 1   \\
				\hline
				2    & 0.1 & -3.14 & 0.9227 &  0.9485 & 0.9209  \\
				\hline
				3    & 0.2 & -17.02 & 0.8403 &  0.8571 & 0.7670   \\
				\hline
				4    & 0.3 & -26.31 & 0.7668 &  0.6975 & 0.5676    \\
				\hline
				5    & 0.4 & -39.35 & 0.7978 &  0.8875 & 0.4647   \\
				\hline
				${\alpha _l}$ & \multicolumn{5}{c}{Lognormal distribution, $LN( - 3.66,1.08)$}   \\
				\hline
				${\phi _l}$    &  \multicolumn{5}{c}{Uniform distribution, $U[ 0,\pi]$}   \\
				\hline
				\makecell[c]{${f_{D,l}}$}  & \multicolumn{5}{c}{Uniform distribution, $U[ -{f_{max}},{f_{max}}]$} \\
				\hline
				\hline
				
			\end{tabular}
			
		\end{threeparttable} 
	\end{center}
\end{table}

\subsection{Model Implementation}
The measured data  serve as the foundation for  model establishment. 
Notely that at least four complete cycles of valid measured data are obtained,
two of which have been  previously used in \cite{zxjwcsp} to investigate large-scale and small-scale propagation characterization, including path loss,  PDP, RMS DS,  etc., but do not involve channel models suitable for link-level simulation.
Here, we use the data from the other two cycles to establish   Markov TDL model and then verify it.
  
Based on measured data, we parameterize the number of taps, the relative tap delay and the  tap power of TDL model as listed in Table \ref{TDL}. 
Parameter estimations are given as follow:

1) Determine the   number of channel taps $L$. 
The number of taps employed in the TDL model is widely computed by the maximum value of  RMS DS divided by the time resolution of the channel sounder \cite{ sen}, then it can be approximated as 
\begin{equation}
	\label{math3}
{L} = \left\lceil {\frac{{\max ({\sigma _{\tau}})}}{{{T_c}}}} \right\rceil  + 1,
\end{equation}
where  $\sigma _{\tau}$ is the RMS DS, and $T_c$ is the coherence time.
In this paper, the maximum value of $\sigma _{\tau}$ is slightly larger than 350 ns, which can be found in \cite{zxjwcsp}.
And  $T_c$ is  100 ns corresponding to 10 MHz bandwidth.
Thus $L$ is determined to be 5.

2) Extract the amplitude $\alpha _l$ and phase ${\phi _l}$ of each tap from normalized average PDPs (APDPs), so that shadow fading and path loss are excluded in the normalization. 
An example of APDP can be seen in \ref{amplitute}(a).
While determining the average energy, only tap samples with relative energy above a threshold (6 dB above the background noise floor \cite{zhangbei,MiYang2023}) are counted.
The measured results and cumulative probability functions
(CDFs) of amplitudes are shown in Fig. \ref{amplitute}(b), and
it can be found that  the Lognormal distribution can fit measured data  well. 
What's more, the phase follows the Uniform distribution over the range of $[0, \pi]$ for all taps.
The above results are consistent with \cite{ans2024, hassan}. 

3) Determine the distribution of Doppler shift. 
Doppler shift is used to describe the time selectivity of   wireless channel and is mainly caused by the relative motion between Rx and Tx \cite{myy}.
The Doppler frequency shift $f_D$ can be theoretically calculated as \cite{liu2012, zy2014} 
\begin{equation}
	\label{math4}
{f_D} = \frac{v}{\lambda }\cos (\theta (t)) = {f_{\max }}\cos (\theta (t)),
\end{equation}
where  $v$ is the velocity in meters per second, $\theta (t)$  is the angle of incidence, and $\lambda$ is the wavelength. 
The maximum speed of Rx in our measurements is 80 km/h, which corresponds to a maximum Doppler frequency shift $f_{\max }$ =160 Hz. 
The Doppler feature of the taps can be modeled as a random variable $f_v$, which follows a Uniform distribution, ${f_v} \sim U[ - {f_{\max }},{f_{\max }}]$ \cite{zy2014,ans2024}.  

4) Assign tap state 0 or state 1 to all extracted taps of each snapshot. 
Apply tap threshold to decide the disappearance or existence of taps, set 0 or 1. 
Calculate tap transition probabilities between state 0(1) to state 1(0), including $p_{00}$, $p_{01}$, $p_{10}$ and $p_{11}$. 
For brevity, only $p_{00}$, $p_{11}$ and $p_1$ are shown in Table \ref{TDL}. 
Here, $p_1$ can be achieved as ${p_1} = {p_0}{p_{01}} + {p_1}{p_{11}}$.
It is observed that the probabilities of $p_{00}$ and $p_{11}$ are dominant, indicating that when an MPC is ``birth" or ``death" at the current moment, it   maintain the current state with a high probability at the next moment.
Furthermore, the probability of $p_1$ applied to the 1st to 5th taps decreases gradually, primarily because a significant portion of   measured data can clearly distinguish only 2 or 3 taps. 
And it implies that at least half of the snapshots cannot resolve 4 or 5 taps due to the limited delay resolution and  sparse  environmental scatterers.

5) Calculate correlation matrix between different taps.
The classic TDL model under the WSSUS assumption does not involve correlations between different taps, but the Markov-based TDL model does. 
It is critical in the design and performance evaluation of advanced signal processing algorithms at  Rx \cite{ans2024, hassan, ha2021}.
The correlation coefficient between two taps is estimated as \cite{hassan}
\begin{equation}
	\label{math5}
{C_{i,j}} = \frac{{{\mathop{\rm cov}} ({\alpha _i}{\alpha _j})}}{{\sqrt {{\mathop{\rm var}} ({\alpha _i}){\mathop{\rm var}} ({\alpha _j})} }},
\end{equation}
where $\alpha _i$ and $\alpha _j$ represent the magnitudes of taps $i$ and $j$.
The terms ${\mathop{\rm cov}} ( \cdot )$ and ${\mathop{\rm var}} ( \cdot )$ denote the covariance and variance functions, respectively.
Table \ref{corr} lists the   correlation coefficients for  five-tap model we proposed, which is symmetric with respect to the diagonal.
Thus we only need to determine the upper triangular part in fact.

\begin{table}
	\renewcommand{\arraystretch}{1.2} 
	\begin{center}
		\caption{Tap Amplitude Correlation Coefficients.}
		\label{corr}
		\begin{threeparttable}
			\begin{tabular}{ c| c| c| c|   c| c }
				\hline
				\hline
				Tap  & 1  &2  & 3 &  4 & 5   \\
				\hline
				1   & 1 & 0.5009 & 0.7733 &  0.2320 & 0.0525   \\
				\hline
				2    & 0.5009 & 1 & 0.6170 &  0.5997 & 0.1714  \\
				\hline
				3    & 0.7733 & 0.6170 & 1 &  0.4369 & 0.0513   \\
				\hline
				4    & 0.2320 & 0.5997 & 0.4369 &  1 & 0.6132    \\
				\hline
				5    & 0.0525 & 0.1714 & 0.0513 & 0.6132 & 1   \\
				\hline
				\hline
				
			\end{tabular}
			
		\end{threeparttable} 
	\end{center}
\end{table}
  
\begin{figure}[!t]
	\centering
	\includegraphics[width=.44\textwidth]{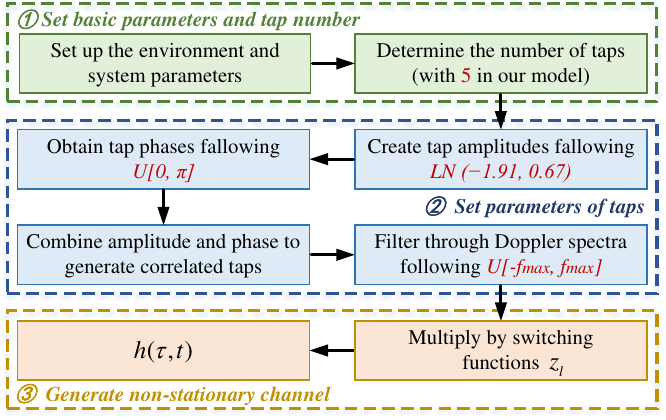}
	\caption{Diagram of Markov-based TDL channel simulation.}
	\label{block}
\end{figure}

\begin{figure}[!t]
	\centering
	\includegraphics[width=.45\textwidth]{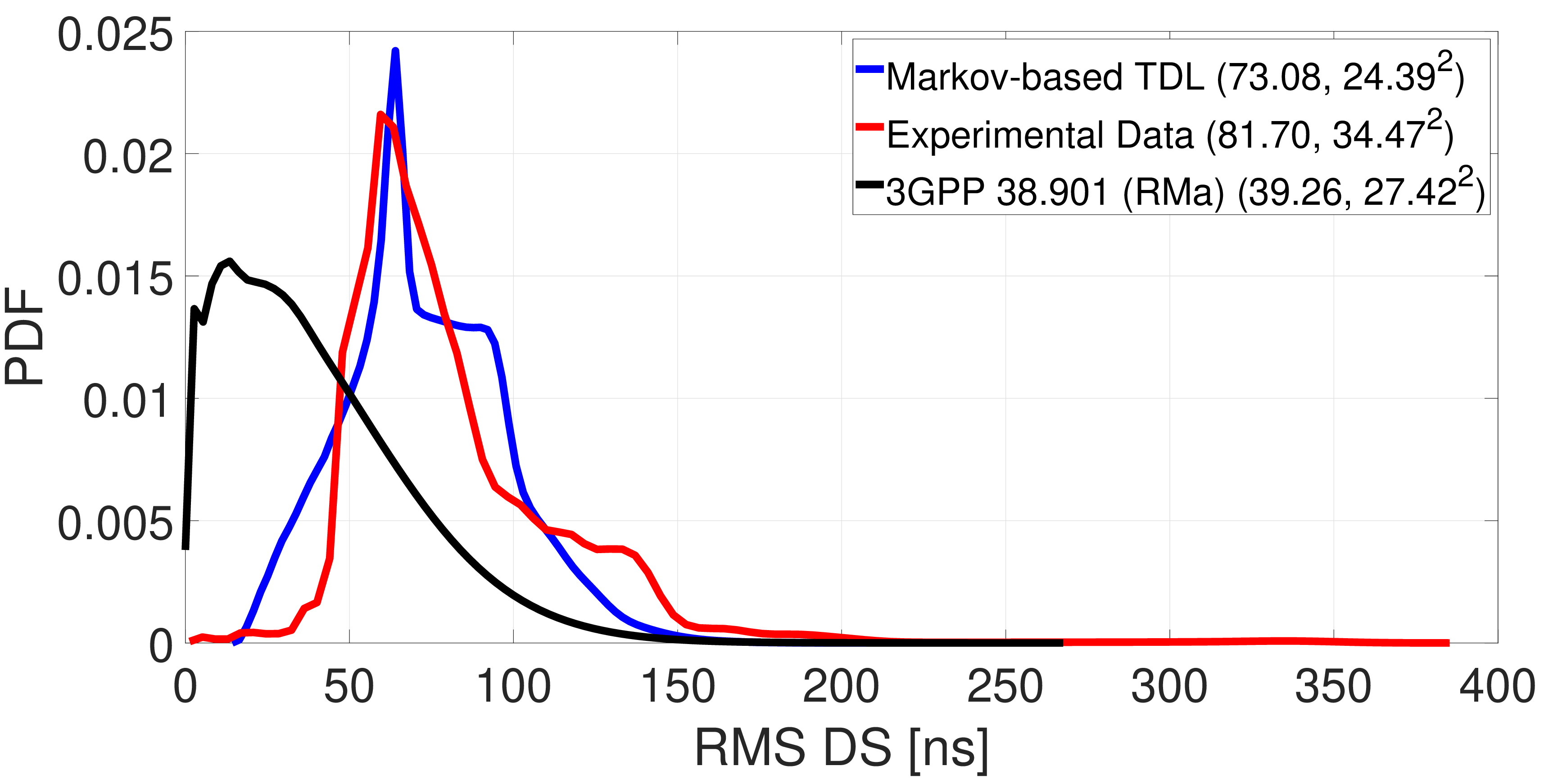}
	\caption{Comparison of normalized RMS DS PDF between Markov-based TDL model, experimental data, and 3GPP 38.901 model (RMa), where $(\mu ,\sigma ^2) $ represents the mean value and standard variation of RMS DS.
	}
	\label{RMS}
\end{figure}

\subsection{Validation}
We use the other group of measured data mentioned in Section III-B to validate the proposed TDL model.
Based on the parameters listed in Table \ref{TDL}, the simulation channel is generated  following three steps  as shown in Fig. \ref{block}:
{\romannumeral1}) Set basic parameters, including measurement scenario, radio frequency
parameters, antenna patterns, etc. 
The number of taps is set to 5, consistent with the result of proposed model.
{\romannumeral2}) Obtain a series of tap parameters based on  statistical distribution of measured data, including amplitude, phase, Doppler shift, etc.
{\romannumeral3}) Add the transition function, represented by the first-order two-state Markov chain, to generate a non-stationary channel.

The probability distribution functions (PDFs) of RMS DS between simulated, measured and 3GPP channels are shown in Fig. \ref{RMS}, which can be used to validate the accuracy of the model.
It can be seen that the overall agreement is fairly good and reasonable.
In contrast, the fitting results of 3GPP channel and measured channel are quite different. 
This may be due to the inability of  3GPP model to capture   ``birth and death" processes and accurately characterize   non-stationarity of   channel. 
What's more, 3GPP model lacks adaptability to railway-specific channels and does not provide a channel model tailored to frequency and bandwidth requirements of 5G-R systems.
The above results  clearly demonstrate that the proposed Markov-based model has improved performances than 3GPP model, making it more consistent with dynamic channel conditions.

\section{CONCLUSION}
In this paper,   a wideband channel measurement campaign is conducted for  5G-R private networks. 
Based on the collected data, a 5-tap TDL model utilizing a first-order two-state Markov chain is proposed. 
This model effectively characterizes   ``birth and death" processes of MPCs in railway communications and represents   non-stationarity of  channel. 
Specifically, parameters of the TDL model, including  number of taps, statistical distributions of amplitude, phase, and Doppler shift are extracted. 
The correlations between  tap amplitudes are also obtained.
The proposed model is evaluated against measurement data and 3GPP TDL model in terms of  RMS DS, validating its fairly good performance, better than  3GPP model.
The above results  can provide necessary guidance for link-level simulation, performance evaluation, and air interface technology optimization of 5G-R systems.

   \balance
   \bibliographystyle{IEEEtran}

   \nocite{*}
   
   \bibliography{IEEEabrv,ref}
\end{document}